\documentclass[manuscript]{aastex}
\begin{document}

\title{Hydrogen lines in LAMOST low-resolution spectra of RR Lyrae stars}

\author{F. Yang\altaffilmark{1,2}, L. Deng\altaffilmark{1} ,  C. Liu\altaffilmark{1}, J. L. Carlin\altaffilmark{3}, H. Jo Newberg\altaffilmark{3}, K. Carrell\altaffilmark{1}, S. Justham\altaffilmark{1}, X. Zhang\altaffilmark{1}, Z. Bai\altaffilmark{1}, F. Wang\altaffilmark{1}, H. Zhang\altaffilmark{1},K. Wang\altaffilmark{1,4}, Y. Xin\altaffilmark{1}, Y. Xu\altaffilmark{1}, S. Gao\altaffilmark{1},  Y. Zhang\altaffilmark{1}, J. Li\altaffilmark{1}, Y Zhao\altaffilmark{1}}
\affil{$^1$Key Lab for Optical Astronomy, National Astronomical Observatories, Chinese Academy of Sciences, Beijing 100012, China}
\affil{$^2$University of Chinese Academy of Sciences, Beijing 100049, China}
%
\affil{$^3$Department of Physics, Applied Physics and Astronomy, Rensselaer Polytechnic Institute, 110 8th Street, Troy, NY 12180, USA}
\affil{$^4$School of Physics and Electronic Information, China West Normal University, Nanchong 637002, China}


\begin{abstract}

The LAMOST pilot survey has produced a data release containing over 
600,000 stellar spectra. By cross-checking with a large time series photometric database of RR 
Lyrae stars in high Galactic latitude regions, we found a total number of 157 RR Lyrae stars that have been observed with LAMOST. In this sample, 
we successfully captured three RR Lyrae stars in the fast expansion phase, all of them showing hypersonic shock wave features in the Balmer line region. 
We fit the shape of H$\alpha$ line region and determine that the emission feature seen within the broadened H$\alpha$ absorption line suggests hypersonic relative motion in the atmospheres of these three objects. With a further LAMOST survey of millions of stars, we plan to capture a large sample of RR Lyrae stars in their hypersonic expansion phase, and therefore provide a large database for the study of the internal structure and the pulsation mechanism of RR Lyrae stars.
\end{abstract}


\section{Introduction}


RR Lyrae stars show variabilities in observational quantities including radial velocity, luminosity and color due to their pulsations. The RR Lyrae pulsation features a slow contraction after a fast expansion, which can be detected using radial velocity line shifts.  
Although generally showing regularity in their pulsation, RR Lyrae stars have very complicated atmospheric hydrodynamic processes that have been revealed by high resolution spectroscopy of a few bright RR Lyrae stars(e.g., Chadid et al. 2008; Chadid 2011). 
Evidence that hypersonic shock wave, a shock wave where radiative phenomena have major a impact, exists in the RR Lyrae stars has been reported by Chadid (2008). Hereafter, in this paper we define the hypersonic shock phase as the phase when the hypersonic shock wave takes place. Observations indicate that the strong shocks propagate for about 4\% of the period in a pulsation cycle, based on the evidence of shock signatures such as line doublings and emission in hydrogen lines (Preston \& Paczynski 1964; Preston et al. 1965). For large amplitude RR Lyraes, motions in the atmosphere can become hypersonic. During a pulsation cycle, minor emission caused by the shock of ballistically falling outer material onto the bottom of the atmosphere during the bump phase (at phase $\sim 0.7$) can be observed in high resolution spectra (Preston 2011). 

Spectral signatures in the fast expansion phase include emission features in the Balmer lines (Preston 2011), the appearance of helium lines, and the disappearance of neutral metallic lines (Chadid \& Gillet 1996; Chadid et al. 2008). In the Balmer lines, a prominent narrow emission line is coupled with a broad, red shifted absorption feature. The emission line is explained by a radiation-dominated shock wave propagating in the outer photosphere (Chadid et al. 2008). Meanwhile, the intensity of the absorption lines becomes weaker as the shock wave pushes outwards, and then becomes strong again when the shock wave is no longer radiation-dominated. 

In order to better understand the physics of RR Lyrae pulsations, spectroscopy of a large sample is needed. Unfortunately, the duration of the hypersonic shock phase is typically around 30 minutes, much shorter than most of the exposure times required by high resolution spectroscopic observations, hence only very few cases have been observationally captured so far (Preston 2011; Preston 2009; Chadid et al. 2008).  

Highly-multiplexed, large sky area surveys such as LEGUE (LAMOST Experiment on Galactic Understanding and Exploration; Deng et al. 2012) offer opportunities to observe a large sample of RR Lyrae stars with medium-low resolution spectroscopy. In this paper we present discoveries of this rare phenomenon by the pilot mission of LEGUE, and discuss another viable way to capture the hypersonic expansion phase in RR Lyrae stars using low resolution multi-epoch spectroscopy.


\section{Data and observations}

The Catalina Survey Data Release 1 (DR1) of time series photometric data has recently been published, including a catalogue containing 12397 type-ab RR Lyraes (RR ab) identified from fitting of the light curves. It covers about 20000 deg$^2$ of sky area with $0^{\circ} < \alpha < 360^{\circ}$ and $-22^{\circ} < \delta < 65^{\circ}$ (Drake et al. 2013). In this catalog, the RR Lyrae stars are essentially uniformly distributed, and the upper limit of their heliocentric distances is about 60 kpc (Drake et al. 2013). The visual magnitude limit of the catalogue goes from 11 to 20 Mag in V band.

The footprint of the LEGUE survey (Deng et al. 2012; see also Chen et al. 2012; Yang et al. 2012; Zhang et al. 2012) largely overlaps with the Catalina catalog. Thus, it gives us an opportunity to study RR Lyraes with LEGUE spectra. The system carrying out the LAMOST survey has 4000 fibers in the focal plane, corresponding to a roughly 20 deg$^2$ field of view (see Cui et al. 2012 for more details). The resolution of LEGUE spectra is about R$\approx$1800 with a spectral wavelength coverage of 3700$<$$\lambda$$<$9100$\rm\AA$ (Zhao et al. 2012). By the end of the survey, nearly 10 million spectra of stars in the Galaxy will be collected. This huge number of spectra should be of great help to better understand the nature of our Galaxy, such as the Galactic merger history and disk substructure and evolution. A pilot survey aimed at testing the system performance was conducted from Oct 2011 to Jun 2012. Each set of targets (a plate) is a combination of 2-3 separate exposures. The duration of each single exposure varies from 600 to 1800 seconds, which is approximately  the typical duration of the RR Lyrae hypersonic shock phase. This cadence plus the sheer number of spectra that LAMOST will observe makes it promising for studying the physics of hypersonic shocks in RR Lyrae stars.
 
More than 600,000 stellar spectra have been released from the LAMOST pilot survey (Luo et al. 2012). Cross-matching the Catalina catalogue with the LEGUE pilot survey data found 157 RRab stars which have been spectroscopically observed by LAMOST. In standard data processing, object spectra are extracted by a 2D pipeline, then combined after wavelength calibration and sky subtraction (Zhao et al. 2012; Guo et al. 2012). Therefore the regular data release does not contain information from single exposures. 
Because RR Lyraes pulsate over short time scales (roughly a few hours), we can use the individual spectra from repeated exposures of the same plate to explore changes in the properties of these stars on $\sim30$-minute time scales.
Instead of combining the single spectra of all exposures of a plate and running regular 1D pipeline tasks, we directly extract the single spectra of all RR Lyrae targets from the 2D pipeline process. These single spectra have already passed flat fielding, wavelength calibration and sky subtraction, and are ready for the analysis described below. In the end, a total number of 397 single-epoch spectra of those 157 stars were collected. Each star has two or three exposures at different epochs, corresponding to two or three spectra.

\section{Analysis and results}
LEGUE covers the whole optical wavelength range using a blue arm (3700-6000~\rm\AA) and red arm 5700-9100~\rm\AA. The system performance is typically better in the red arm in terms of throughput and wavelength calibration. To ensure clarity and uniform velocity measurements, 
we use only the H$\alpha$ line in the red arm as the radial velocity indicator, and only locally normalise the spectra in the range of 6450$<\lambda<$6750~\rm\AA. The quantity $\sigma$ for each of the spectra is then calculated from the RMS of the two fractions of the spectra at  6450$<\lambda<$6500~\rm\AA~ and 6700$<\lambda<$6750~\rm\AA, within which the spectra are continuum dominated. The signal to noise ratio (S/N) is obtained by S/N = 1/$\sigma$. We select objects with S/N $>15$ for the following analysis, leaving a sample of 211 spectra belonging to 99 individual stars. 

\subsection{Search for stars observed during the fast expansion phase}
For the purpose of the current work, it is not necessary to measure the radial velocity from each single spectrum. The differential velocity between two epochs is sufficient for the phase determination of the RR Lyrae stars. Hence, we apply a cross-correlation method to estimate the differential radial velocity, $\Delta$(RV), of each pair of spectra for a given star. To satisfy this requirement, each star must have more than one spectrum among those 211 high S/N spectra, which gives us a final sample of 184 spectra from 72 stars. Among those stars, 40 stars have three spectra and 32 stars have two spectra. Finally, we have 152 pairs of spectra with corresponding differential radial velocities from 72 stars in total.      

In order to obtain the velocity shift for each pair of spectra, the first observed spectrum in a pair is treated as a template and the second one is shifted by 1 km~s$^{-1}$ steps over a range of -250 km~s$^{-1}$ to 250 km~s$^{-1}$. The cross-correlation index of the two spectra at each step is calculated from the normalised spectra within 6510$<\lambda<$6610~\rm\AA, in which the H$\alpha$ line is located. The $\Delta$(RV) is determined as the velocity shift at which the index peaks. This is regarded as the most likely $\Delta$(RV) of the pair.

Although the Catalina catalogue provides several observable quantities for the RR Lyrae stars, including the pulsation period and amplitude, the ephemeris given in the catalogue is not reliable now because the observations were taken more than 7 years ago.  Despite this, we can still simulate the distribution of differential radial velocity for LEGUE observations for a given period and normalised amplitude.  For each of the 72 stars in our sample, we first randomly select a zero point phase between 0 and 1 for the initial exposure.  Then, the simulated phase of the second and third (if it exists) exposure is determined from the temporal spacing of the observations and the known pulsation period from the Catalina catalog.  The $V-$band light curve amplitude of each star's variability (in magnitudes) from the Catalina catalogue is converted to a radial velocity amplitude using the relations for these quantities from Sesar (2012). Then the phase-RV$_{norm}$ relation given by Sesar (2012) is used to derive the radial velocity of each simulated exposure.  The phase-RV$_{norm}$ relation for the Balmer lines given by Sesar (2012) is not valid for phase $>$ 0.95, and therefore a linear interpolation is adopted for this region. The observations for each star are simulated 500 times. We then bin the differences between the radial velocities of the initial and subsequent phases. A histogram of the results for these 76,000 simulated pairs of spectra (scaled to match the total of 152 observational points) is shown in Figure~1 (lower panel). Theoretically, every point with a negative value directly derived from this method means at least one of the spectra in the pair was taken during the expansion phase. 
However, we must also consider the error in the observed radial velocities. We thus convolve the simulated distribution with a Gaussian kernel of width 10~km~s$^{-1}$, which is roughly the error in the radial velocities with this observational setup. Stars with large negative $\Delta$(RV) values making up the long thin tail in Figure~1 can be considered to show strong evidence that the spectra were taken in the expansion phase.
In the distribution of observed differential radial velocity (the upper panel of Figure~1), there is a sudden drop around $\Delta$(RV) $\sim$ $-30$~km~s$^{-1}$. We select stars with $\Delta$(RV) lower than $-30$~km~s$^{-1}$ as candidates which may have been observed during the fast expansion phase. There are 13 pairs of spectra from 9 stars that show values lower than $-30$~km~s$^{-1}$. From the simulation we predict that for an observation with 152 pairs of spectra, we will get around 11 pairs having $\Delta$(RV) lower than $-30$~km~s$^{-1}$, which agrees very well with the real observation.
%

\subsection{Examination of fast-expanding stars for hypersonic shock signatures}
We visually investigated the spectra of stars that were selected as fast expansion phase candidates.
Three of the objects were found to show clear Balmer emission line features together with shallower absorption lines in one of their spectra (the thin red lines in Figure~2), which indicates that these spectra were taken during the hypersonic shock phase. For comparison, the other spectrum of the same star taken outside the fast expansion phase (the thick blue lines in Figure~2) shows a stronger Balmer absorption profile typically observed in RR Lyrae spectra. These three objects are indicated in the distribution histogram in Figure~1 as bins with light shading and some of their parameters are given in Table 1. Because in the cross-correlation method we assume the shape of the H$\alpha$ feature is an absorption line, an overlapped and blue shifted emission line may result in an extra negative $\Delta$(RV). This may explain why we have an even lower differential radial velocity than the simulation for the star LRR1 ($\sim150$~km~s$^{-1}$). Thus by looking at exceptionally low differential radial velocities acquired from the cross-correlation method, there is a high probability that we can find Balmer emission line features in the RR Lyrae spectra.
Figure~2 shows the Balmer line profiles in the spectra of these three stars. Because of the different beginning epochs of the two exposures (see Table 1), the thin lines are likely taken in the fast expansion phase, while the thick lines are during the phase of slow expansion. The emission components can be clearly identified in multiple line profiles for those three objects.

To check whether the apparent Balmer emission line is caused by over-subtraction of the sky lines, we examine the emission lines in each sky spectrum individually. The nearby sky lines are at least $1-2~\rm\AA$ offset from the H$\alpha$ emission lines in the stellar spectra. The nearest sky line (around 6564~\AA) is only about 50 units (ADU) higher than the continuum of the sky spectra, while the H$\alpha$ emission line is about 300 units higher than the continuum in the stellar spectra. Therefore, the emission in H$\alpha$ cannot be the result of poor sky subtraction. More importantly, not only does H$\alpha$ show an emission line, but also H$\beta$ and H$\gamma$ (which are far less affected by the sky lines) independently show clear emission features as well. The overall weakening of the Balmer line profiles in the expansion phase is also seen in all three stars. There is no sign of emission lines in the Paschen lines, which agrees with the behavior seen by Chadid (2011). 
Between the spectra with and without emission lines, we do not see significant differences in the He lines (Preston 2009). The CaII H and K lines do have differences, but these could be related to the different effective temperatures between the two phases rather than the hypersonic shock wave directly.  
We were unable to visually identify significant differences between the strengths of neutral metallic lines in and out of the fast expansion phase, as have been seen at high resolutions (e.g., Chadid et al. 2008). This is likely due to a combination of both low spectral resolution and modest S/N in the LAMOST spectra.
Finally, we note that in all three cases, the emission spectrum has more and stronger ``wiggles'' which may also be due to a lower S/N and/or a cooler temperature for the star during this phase. 
We have made a recent photometric observation dedicated to ephemeris and period updates, which confirmed that the LRR1 spectrum with emission features was
indeed taken at a phase of 0.9$\sim$0.94. More follow-up observations will be
carried out with a detailed analysis available in a forthcoming paper. 

We now estimate whether this number of detections of the hypersonic shock phase is reasonable within our sample. Assuming an exposure time of 10\% of the pulsation period and a fast expansion phase that lasts 5\% of the period, we estimate an upper limit on the probability of detecting the fast expansion phase of 15\%.
According to this, we expect to find $\sim11$ RR Lyrae in the fast expansion phase among the 72 RR Lyrae stars found in the LEGUE pilot survey. Indeed, we find 9 stars with $\Delta$(RV) $< -30$~km~s$^{-1}$, which are likely in the fast expansion stage. Because only RR Lyrae with high amplitude can trigger the hypersonic shock wave, not all of these 9 candidates show Balmer emission lines. If we assume the RR Lyrae stars with amplitude higher than 0.9 mag in V band can excite the Balmer emission lines, there should be only three stars in our sample having such features. This is exactly the number of stars we find. Furthermore, these particular stars do display high amplitude pulsations (see Table 1).  Therefore, the discovery reported in this paper is highly in agreement with the radial velocity - period relationship of RR Lyrae stars. 

\subsection{Analysis of spectra taken during hypersonic shock phase}
We next estimated the shock front velocity by finding the relative velocity of the blue shifted emission line and the red shifted absorption line. The emission feature is associated with the shock itself, whilst the absorption feature comes from the outer parts of the star, and thus moves with the bulk expansion of the star.
To determine the relative velocities of these two features, we fit the H$\alpha$ region with a narrow emission line superimposed on top of an absorption line. S\'ersic profiles are applied to fit both lines (Xue et al. 2008). These fits are shown in Figure 3 for the candidate fast-expansion phase RR Lyrae. 
The measured velocity differences between the two components of the H$\alpha$ lines are 195, 107 and 62 km~s$^{-1}$.

Chadid et al. (2008) found that RR Lyrae shock velocities of 70 ~km~s$^{-1}$ correspond to shock Mach numbers $\sim$ 10, and therefore correspond to radiation-dominated hypersonic shocks. We therefore conclude that the shock waves we found in those 3 RR Lyrae stars were hypersonic and those 3 RR Lyrae stars were indeed observed during the hypersonic shock phase with extremely high contrast between the motions of different layers.

For LRR3, we infer the velocity of the shock wave front to be about 62 ~km~s$^{-1}$, which suggests that this could be a particularly interesting case. We apply Rankine-Hugoniot relationships to estimate the temperature of the shock wave  (Chadid et al 2008) and the value is about 71650 ~K. This temperature is able to ionise a neutral atom of Ca but unable to ionise neutral atoms of Fe, Ti and Mg (Chadid et al 2008). Assuming that 62 ~km~s$^{-1}$ is an accurate measurement of the shock velocity in LRR3, we would expect to see the neutral metallic lines to partially disappear with the only disappearance of neutral Ca lines during the hypersonic shock phase in this system. Observing such a partial disappearance of the neutral lines would enable us to study the temperature behind the shock by using the different ionisation potentials of metallic species as a probe. This would require a high resolution spectroscopic observation of LRR3, ideally with shorter exposures. One potential systematic effect affecting our 62 ~km~s$^{-1}$ measurement is our 30 minutes exposure. The real velocity may be even higher, as would be the corresponding shock temperature.


 

\section{Conclusions}

Using the separate, single exposures from consecutive spectroscopic observations of the same targets, the LEGUE survey provides a time-domain capability to spectroscopically study the variability of properties of stars. In the LAMOST pilot survey data set, three RR Lyrae stars are discovered that show hypersonic expansion features in their low resolution spectra. The differential motions of the atmospheric layers detected by the emission and absorption components of H$\alpha$ for the 3 stars are 195, 107 and 62 km~s$^{-1}$.  This strongly indicates hypersonic motions in the atmospheres of these stars. LRR3 may enable measurements of the temperature behind the shock front if future high resolution spectroscopic observation confirms partial disappearance of the neutral metallic lines during the hypersonic shock phase. 

The cross-correlation method can be applied to future survey data, which can efficiently capture this quite short ($\sim$ 5$\%$ of the pulsation period) yet very important phase of RR Lyrae stars, especially the hypersonic shock wave phenomenon. 

Apart from discovering hypersonic shock waves in the RR Lyrae stars, the LEGUE spectra can also help to determine their intrinsic radial velocities. The data shows that many objects are observed three times within 1-2 hours in the LAMOST survey. For RR Lyrae stars, these three single exposures will provide three radial velocities, which includes the intrinsic radial velocity of the star and the additional positive (in contracting phase) or negative (in expanding phase) pulsating velocities, at different phases. Given the amplitude and period of an observed RR Lyrae star from external sources, e.g.~the Catalina catalog, the phase-velocity relation of the star can be rescaled from a standard template (Sesar 2012) to fit the observation so that the pulsating velocity can be removed and the intrinsic radial velocity of the star can be determined.

Using the methods outlined in this work, the millions of spectra to be obtained by the LEGUE survey will yield the following results on RR Lyrae stars:

\begin{enumerate}
\item By using a large sample of single exposure spectra from LEGUE, we can discover RR Lyrae stars by statistical analysis of the differential radial velocity distribution function, with or without emission features in one of three single exposures.
\item We will capture a large spectroscopic sample of RR Lyrae stars with hypersonic shocks and create a large database for the study of RR Lyrae pulsations.
\item The intrinsic radial velocities of RR Lyrae stars can be derived by analysing individual LAMOST spectra of objects with repeat observations. RR Lyrae stars are ideal tracers for the structure of the Milky Way halo because they are intrinsically bright and their distances can be accurately determined. When searching for kinematic substructure in the halo, it is important to derive intrinsic radial velocities from multiple exposures of each star in order to account for the RV variability due to pulsations (especially for those that are caught in the fast expansion phase).

\end{enumerate}

\acknowledgements{
Guoshoujing Telescope (the Large Sky Area Multi-Object Fiber Spectroscopic
Telescope LAMOST) is a National Major Scientific Project built by the
Chinese Academy of Sciences. Funding for the project has been provided by
the National Development and Reform Commission. LAMOST is operated and
managed by the National Astronomical Observatories, Chinese Academy of
Sciences.  This is supported in part by NSFC through grants 10973015. CL is supported by NSFC grant U1231119. YX is supported by NSFC grant Y111221001. JLC and HJN are supported by NSF grant AST 09-37523.}

\begin{figure}
\epsscale{.75}
\plotone{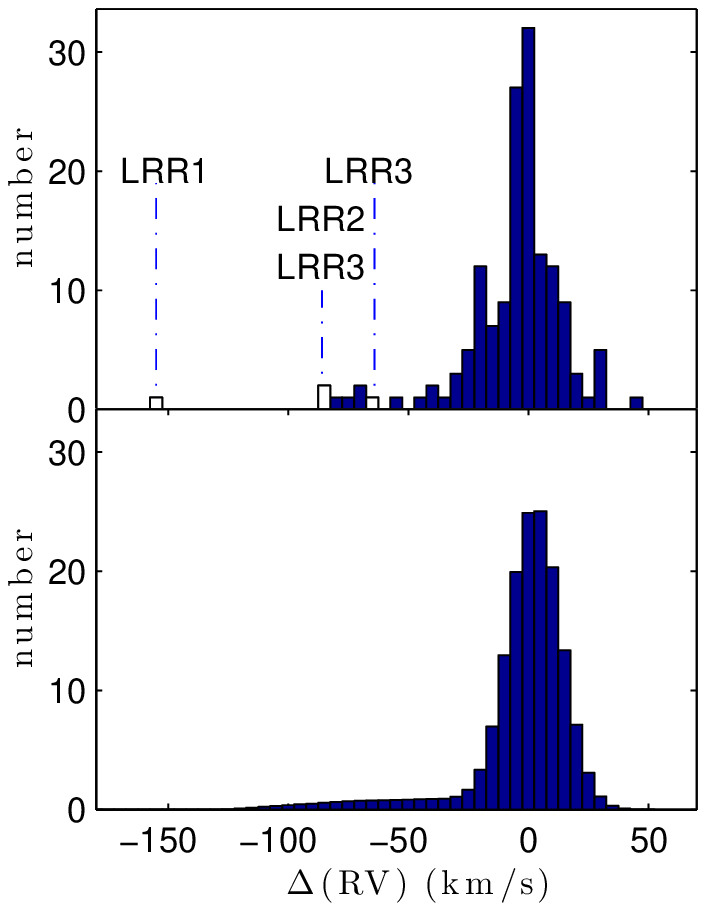}
\caption{Upper panel: Observed differential radial velocity for all pairs of spectra. The differential radial velocities from four pairs of spectra of three objects with significant emission features are shown as white bins. Star LRR3 has two out of three pairs of differential RV less than $-50$~km~s$^{-1}$. Lower panel: Differential radial velocities from the simulation described in Section~3. The total number has been normalised to 152 to compare with the observed result. The positive values imply contraction while the long tail of negative values mean the fast expansion phase of RR Lyraes is captured.}
\end{figure}

\begin{figure*}
\epsscale{1.0}
\plotone{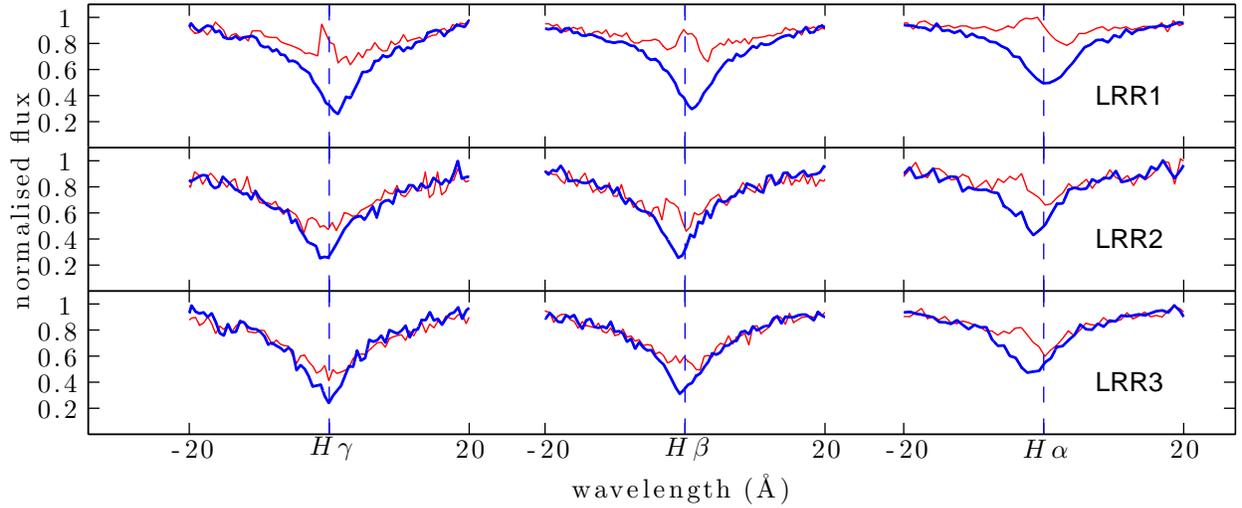}
\caption{Balmer lines for three RR Lyrae stars (spectrum pairs) with differential radial velocity lower than $-50$~km~s$^{-1}$. The thin red line is the first spectrum, while the thick blue one is the second spectrum. All of the firstly taken spectra show clearly blue shifted emission components within the broader absorption feature of the H$\alpha$, H$\beta$, and (to a lesser extent) H$\gamma$ lines.  These blue-shifted emission features imply the existence of hypersonic shock waves in these RR Lyrae. 
  }
\end{figure*}

\begin{figure}
\epsscale{.9}
\plotone{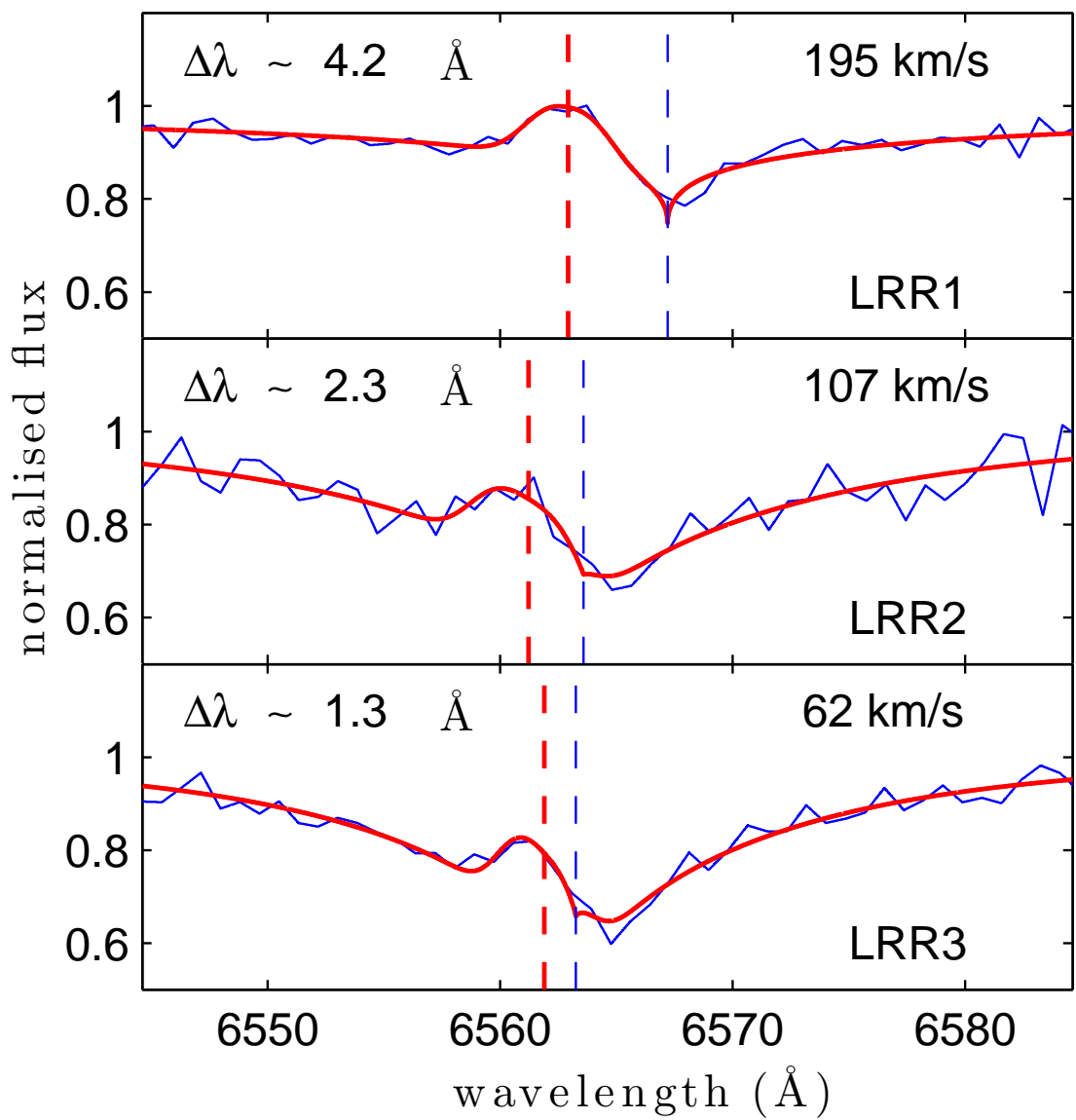}
\caption{Results of fitting an emission line superimposed on top of an absorption feature to the observed H$\alpha$ line profile of the three stars with fast expansion signatures. The thin blue solid line is the observed spectrum while the thick red solid line is the fit. The thick red dashed vertical line shows the emission line center and the thin blue dashed vertical line shows the absorption line center. The differences between the line centers of the two components in both wavelength and RV are indicated in each panel. 
}
\end{figure}

\begin{table*}
\begin{center}
\caption{RR Lyrae stars in the LEGUE pilot survey with hypersonic shock expansion.\label{tbl-1}}
\begin{tabular}{crrrrrcrcrcrrrrr}
\tableline\tableline
star\tablenotemark{a} & R.A. & Dec. & Vmag & A\tablenotemark{b} & P\tablenotemark{c} & E1\tablenotemark{d} & D1\tablenotemark{e} & E2\tablenotemark{d} & D2\tablenotemark{e}& $\Delta$(RV)\tablenotemark{f} & V$_{shock}$\tablenotemark{g} \\
\tableline
LRR1 &144.45 & 0.16 & 14.67 & 0.99 & 0.55 & 80602525 & 30 & 80602583 & 30 & -154 & 195\\
LRR2 & 139.79 & 26.96 & 15.41 & 1.04 & 0.53 & 80463189 & 10 & 80463206 & 10 & -82 & 107\\
LRR3 & 245.90 & 36.42 & 14.38 & 1.12 & 0.57 & 80638838 & 15 & 80638883 & 15 & -84 &  62\\
\tableline
\end{tabular}
\end{center}
\tablenotemark{a}{ID of RR Lyrae stars in LEGUE}
\tablenotemark{b}{Pulsation amplitude (in magnitudes)}
\tablenotemark{c}{Period (in days)}
\tablenotemark{d}{Epoch of the beginning of the exposure (Julian day converted to minutes)}
\tablenotemark{e}{Duration of the exposure (in minutes)}
\tablenotemark{f}{Differential radial velocity between the two exposures (km~s$^{-1}$)}
\tablenotemark{g}{Shock velocity inferred from spectral fitting (km~s$^{-1}$)}
\end{table*}

\end{document}